%% file: main.tex
\documentclass[sigconf]{acmart} 

\AtBeginDocument{\providecommand\BibTeX{{%
    \normalfont B\kern-0.5em{\scshape i\kern-0.25em b}\kern-0.8em\TeX}}}

\makeatletter
\def\@ACM@copyright@check@cc{}
\makeatother

\copyrightyear{2025}
\acmYear{2025}
\setcopyright{cc}
\setcctype{by}
\acmConference[UIST '25]{The 38th Annual ACM Symposium on User Interface
Software and Technology}{September 28-October 1, 2025}{Busan, Republic of
Korea}
\acmBooktitle{The 38th Annual ACM Symposium on User Interface Software and
Technology (UIST '25), September 28-October 1, 2025, Busan, Republic of
Korea}
\acmDOI{10.1145/3746059.3747653}
\acmISBN{979-8-4007-2037-6/2025/09}

\usepackage{xspace}\usepackage{graphicx}\usepackage{subfigure}
\usepackage{enumitem}\usepackage{gensymb}
\usepackage{caption}

\setlist[itemize]{leftmargin=*}
\setlist[enumerate]{leftmargin=*,label=\arabic*.}

\newcommand{\name}{\textsc{SightWarp}\xspace}

\newcommand{\GH}{\textsc{GazeToHand}\xspace}
\newcommand{\HG}{\textsc{HandToGaze}\xspace}
\newcommand{\GP}{\textsc{Gaze+Pinch}\xspace}
\newcommand{\TCT}{\textsc{Trial Completion Time}\xspace}
\newcommand{\AT}{\textsc{Acquisition Time}\xspace}
\newcommand{\FMD}{\textsc{First Manipulation Duration}\xspace}
\newcommand{\CC}{\textsc{Clutch Count}\xspace}
\newcommand{\FGC}{\textsc{Failed Gesture Count}\xspace}
\newcommand{\RDM}{\textsc{Hand Rotation}\xspace}
\newcommand{\TDM}{\textsc{Hand Translation}\xspace}
\newcommand{\TECH}{\textsc{Technique}\xspace}
\newcommand{\RM}{\textsc{Rotation Magnitude}\xspace}
\newcommand{\OS}{\textsc{Object Size}\xspace}

\newcommand{\revised}[1]{{\color{blue}#1}}
\renewcommand{\revised}[1]{#1}

\begin{document}

 \title{At a Glance to Your Fingertips: Enabling Direct Manipulation of Distant Objects Through \textit{\name}}



\author{Yang Liu}
\orcid{0009-0005-9839-0864}
\affiliation{%
    \institution{Aarhus University}
    \city{Aarhus}
    \country{Denmark}
}
\email{yangliu.hci@gmail.com}

\author{Thorbjørn Mikkelsen}
\orcid{0009-0003-3753-4153}
\affiliation{%
    \institution{Aarhus University}
    \city{Aarhus}
    \country{Denmark}
}
\email{thormik@cs.au.dk}

\author{Zehai Liu}
\orcid{0009-0007-8695-5041}
\affiliation{%
    \institution{Aarhus University}
    \city{Aarhus}
    \country{Denmark}
}
\email{202303481@post.au.dk}

\author{Gengchen Tian}
\orcid{0009-0004-6277-1422}
\affiliation{%
    \institution{Aarhus University}
    \city{Aarhus}
    \country{Denmark}
}
\email{202303478@post.au.dk}

\author{Diako Mardanbegi}
\orcid{0000-0002-1976-601X}
\affiliation{%
    \institution{American University of Beirut}
    \city{Beirut}
    \country{Lebanon}
}
\email{diako.mardanbegi@aub.edu.lb}

\author{Qiushi Zhou}
\orcid{0000-0001-6880-6594}
\affiliation{%
    \institution{Aarhus University}
    \city{Aarhus}
    \country{Denmark}
}
\email{qiushi.zhou@cs.au.dk}

\author{Hans Gellersen}
\orcid{0000-0003-2233-2121}
\affiliation{%
  \institution{Lancaster University}
  \city{Lancaster}
  \country{U.K.}
}
\affiliation{%
  \institution{Aarhus University}
  \city{Aarhus}
  \country{Denmark}
}
\email{h.gellersen@lancaster.ac.uk}

\author{Ken Pfeuffer}
\orcid{0000-0002-5870-1120}
\affiliation{%
  \institution{Aarhus University}
  \city{Aarhus}
  \country{Denmark}
 }
\email{ken@cs.au.dk}


\renewcommand{\shortauthors}{Liu et al.}

\begin{abstract}
In 3D user interfaces, \revised{reaching out to grab and manipulate something works great until it is out of reach. Indirect techniques like gaze and pinch offer an alternative for distant interaction, but do not provide the same immediacy or proprioceptive feedback as direct gestures.
To support direct gestures for faraway objects, we introduce \name: an interaction technique that exploits eye-hand coordination to seamlessly summon object proxies to the user’s fingertips.
The idea is that} after looking at a distant object, users either shift their gaze to the hand or move their hand into view—triggering the creation of a scaled near-space proxy of the object and its surrounding context. The proxy remains active until the eye–hand pattern is released. 
The key benefit is that users always have an option to immediately operate on the \revised{distant} object through a natural, direct hand gesture. 
Through a user study of a 3D object docking task, we show that users can easily employ \name, and that subsequent direct manipulation improves performance over gaze and pinch. Application examples illustrate its utility for 6DOF manipulation, overview-and-detail \revised{navigation}, and world-in-miniature interaction. Our work contributes to expressive and flexible object interactions across near and far spaces.

\end{abstract}

\ccsdesc[500]{Human-centered computing~Empirical studies in HCI}
\ccsdesc[500]{Human-centered computing~Pointing}
\ccsdesc[500]{Human-centered computing~Gestural input}
\ccsdesc[500]{Human-centered computing~User studies}

\keywords{Input techniques, extended reality, eye-tracking, gaze interaction}



\maketitle

\captionsetup{aboveskip=4pt, belowskip=-4pt}
\begin{figure}[!h]
    \centering
    \includegraphics[width=1\linewidth]{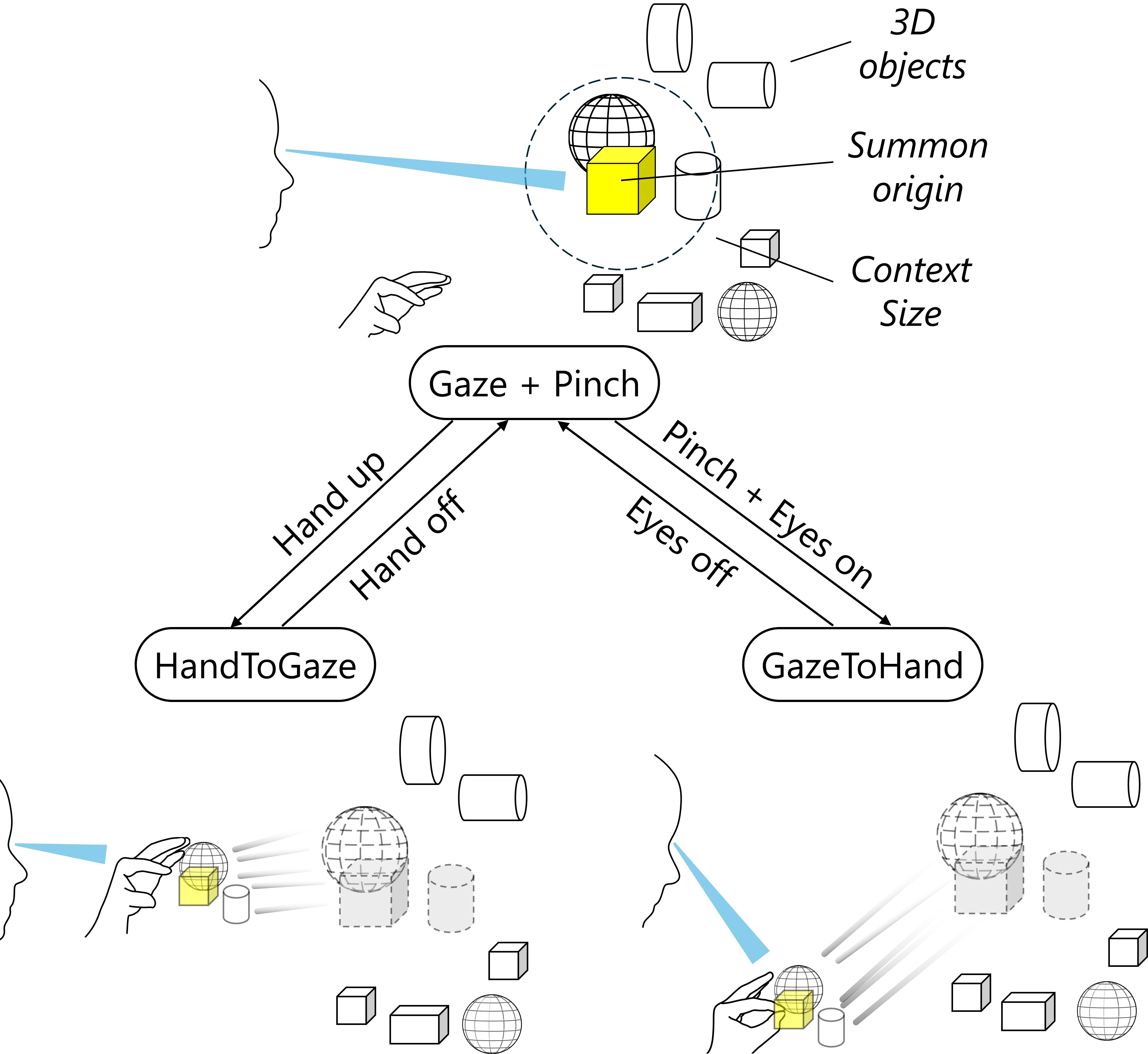}
    \caption{Direct physical manipulation of virtual objects is intuitive. To extend its use across spaces, \name warps distant objects into the user’s hand. By default (top), users manipulate objects at a distance using Gaze+Pinch. Alternatively, they can either move their hand up into their line of sight (left) or shift their gaze down to their hand (right), triggering a proxy that enables direct manipulation of the distant object. This tri-state model allows users to choose the most preferred mode for each gesture.}
    \label{teaser}
\end{figure}

\newpage
\section{Introduction}

The interaction capabilities of extended reality (XR) head-worn computers--experienced by people through headsets and smart glasses--are rapidly evolving. Modern devices, for instance, support hybrid techniques that combine direct and indirect 3D hand-tracking gestures within the same user interface to benefit from the complementary strengths of both input types \cite{hinckley07}.
\textit{Direct gestures} enables physics-oriented interaction with high angular precision, making them compelling for spatial manipulation. When a user grabs a virtual object with their hand, the object moves and rotates in immediate response—creating a tightly coupled feedback loop between motor actions and object behavior. This interaction supports proprioceptive awareness and provides rich spatial cues \cite{mine1997moving}. 
With the integration of eye-tracking, \textit{indirect gestures} allow users to point using their gaze, combined with a pinch gesture for object manipulation ("\GP" \cite{Pfeuffer17,mutasim2021pinch,wagner2023fitts}). The gesturing hand can remain in a comfortable position, reducing physical effort and avoiding hand occlusion of the field of view \cite{lystbaek2024hands,Forlines07}.


%

A key quality of XR interfaces is the support for interaction across depth, from near to far space \cite{mine95}. \revised{In near space, users can fluidly switch between direct and indirect gestures. For far space, techniques such as World-In-Miniatures \cite{WIM}, VoodooDolls \cite{VoodooDolls_Pierce_Stearns_Pausch_1999} and Scaled-World-Grab \cite{mine1997moving} as well as more recent approaches \cite{Poros_Pohl_Lilija_McIntosh_Hornbæk_2021,Xia18} allow users to summon proxies of faraway objects to near space. These methods enable the benefits of direct manipulation through a switching mechanism, but often require a separate invocation gesture \cite{Poros_Pohl_Lilija_McIntosh_Hornbæk_2021} or override the existing indirect interaction mode \cite{mine1997moving, VoodooDolls_Pierce_Stearns_Pausch_1999}. In this research, we focus on extending the default gaze-and-pinch UI semantics with a proxy summoning-- but without asking users to learn extra gestures or give up indirect control. The goal is to make switching between interaction styles feel as seamlessly as in near space.}

We propose \textbf{\name}, a technique that exploits eye-hand coordination for summoning near-field proxies of distant objects in XR UIs. \name is always available, complements \GP, and can be accessed on demand for each gesture. Users begin by identifying a distant object or region of interest using gaze. Then, they can transition into direct gestural manipulation by summoning a proxy of the object and its surrounding context into near space, at the location of their fingertips. This transition is triggered by coordinating gaze and hand in two ways (\autoref{teaser}):

\begin{itemize}
    \item \textbf{\GH}: After initiating a \GP command and holding the pinch gesture, the user directs gaze to their hand. This triggers a context warp, bringing a proxy of the selected object and its local context to the hand, where it appears from a different perspective and is ready for direct gestural manipulation.
    \item \textbf{\HG}: Alternatively, the user can raise their hand into view while maintaining gaze on the distant object. Since users typically do not look at their hand while manipulating distant objects with indirect gestures, this distinct state—gaze focuses on a distant object, with the hand intruding into view—summons the proxy to the hand’s location.
\end{itemize}

\begin{figure*}
    \centering
    \includegraphics[width=1\linewidth]{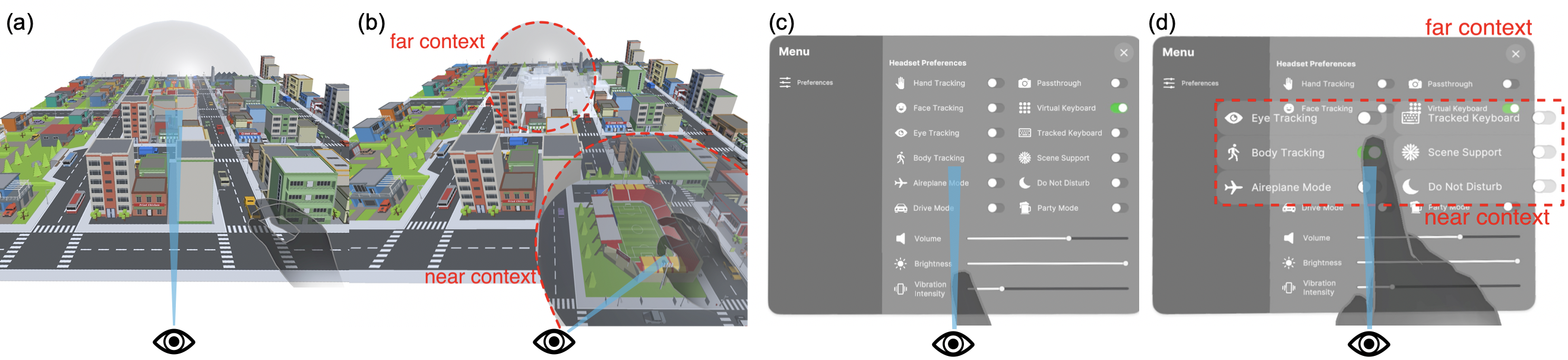}
    \caption{\name applications: (a-b) overview and detail in a city planning tool where users leverage \GH to obtain new perspectives to manipulate detailed or occluded objects; (c-d) \HG for a magnified view of any distant UI by summoning the gaze-focused region to the same fixation distance as the hand and interacting using an air-tap. The eye icon and blue triangle indicate the user's gaze.}
    \label{appteaser}
\end{figure*}

\revised{\name can be useful for various applications}. \GH provides a new perspective by anchoring the summoned object to the user’s hand (\autoref{appteaser}a--b). For instance, when interacting with a city model, \GH enables users to select a specific area and view it from above—facilitating occlusion management and fine-grained object placement in the summoned sub-scene. In contrast, \HG warps content in the user’s forward view closer, functioning like a zoom lens (\autoref{appteaser}c--d). This is particularly useful for distant menus or UI subregions, allowing users to summon a proxy for detailed inspection and interaction, even before deciding what to select or manipulate.

To investigate how effectively users can perform summoning transition and the tangible benefits of proxy-based manipulation, we conducted a user study. \GP was the baseline indirect manipulation technique for distant objects, and we compared it to \revised{our two proposed methods, which employ distinct summoning mechanisms--\GH and \HG--to enable direct gestural manipulation}. We selected a 3D translate+rotate docking task as a standardized, controlled method for gaining insights in user performance and experience. 
The study results indicate the following findings:
\begin{itemize} 
\item Both \name techniques significantly reduced task completion time compared to the baseline, with performance advantages becoming more pronounced as task complexity increased.
\item  \GP led to more clutches and erroneous gestures, reflecting late-trigger issues \cite{Kumar08} and limited preshaping \cite{lystbaek2024hands}.
    \item  \GH resulted in significantly reduced hand movement than \GP, suggesting that the relaxed hand posture facilitates more efficient manipulation.
\end{itemize}

In sum, our results indicate that \name offers measurable benefits over the state-of-the-art eye-hand indirect manipulation technique (\GP) in 3D docking tasks. Notably, the \GH summoning mechanism did not incur performance penalties, suggesting a fluid, low-cost transition into direct manipulation. This makes \name particularly suited for complex spatial tasks such as object docking, alignment, and manipulation--common in 3D design, modeling, and similar domains. Moreover, since \name triggers summoning only through specific eye-hand coordination patterns, it is compatible with existing indirect gestures, allowing XR UIs to support the entire spectrum of direct and indirect interaction across near and far spaces.

The main contributions of this paper to HCI are:
\begin{itemize}
    \item \name, an XR interaction technique that (1) integrates with the existing \GP paradigm, though (2) enabling summoning of remote object proxies with two eye-hand coordination patterns:
\begin{itemize}
    \item \GH: establishes a spatially-distinct perspective as a new context perspective is presented near the hand's location;
    \item \HG: establishes a spatially-consistent perspective as the currently-viewed context is warped to near space;
\end{itemize}
  \item A user study comparing \GH and \HG with the baseline \GP for object docking, showing \name users are more efficient w.r.t. task completion time, clutches, and errors, revealing traits of \GH and \HG.
    \item A set of application examples, demonstrating the utility of \name for cross-space object transfer, occluded and small object selection, details on-demand, and focus-and-context scenarios.
\end{itemize}


\section{Related Work}
We structured the discussion of related work into four parts, described in the following subsections.

\subsection{Distant Interaction in XR}

XR devices incorporate advances in hand-tracking technology that enable users to perform direct 3D manipulation of virtual objects without relying on controllers \cite{mendes2019survey}, fostering a deeper sense of presence \cite{bowman2001introduction} while improving comfort and immersion \cite{hayatpur2019plane}. However, direct hand interaction faces two primary challenges: (1) difficulty interacting with distant or unreachable targets, and (2) fatigue induced by extended hand or arm elevation \cite{hincapie2014consumed, adhanom2023eye,bostan2017hands, siddhpuria2017exploring}. While hand ray-casting 
\cite{wachs2011vision, poupyrev1998egocentric, kyto2018pinpointing,steinicke2006object, gabel2023redirecting} 
provides an effective workaround for the distance issue, studies indicate that hand-based pointing and selection can increase fatigue \cite{Lystbaek22, wagner2024gaze, wagner2023fitts, bowman1997evaluation,yu2020engaging}. Because hands must handle a variety of tasks including pointing, selection via gestures, and subsequent manipulation, their risk of being overburdened further complicates the user experience. 

Gaze interaction offers natural, intuitive, and efficient means of conveying user intent when selecting or interacting with distant virtual objects \cite{tanriverdi2000interacting, jacoby1994gestural}, which has led to extensive research in XR contexts \cite{jankowski2013survey, billinghurst2015survey,Plopski22}. Most XR systems adopt multimodal approaches that combine gaze with other inputs to avoid visual overload from using gaze alone. Common combinations include gaze with head movements \cite{Mardanbegi19, piumsomboon2017exploring, sidenmark2019eye} or hand gestures \cite{Pfeuffer17, Lystbaek22, mutasim2021pinch, wagner2024gaze}. A widely explored division of labour is the “eyes select, hands manipulate” paradigm of \GP \cite{Pfeuffer17}. Studies have shown this approach improves performance for selection compared to hand-raycasting, image plane techniques \cite{wagner2023fitts,Sasalovici25}, and gaze-only methods \cite{mutasim2021pinch, Stellmach12}.

However, the benefits of gaze-hand techniques seem diminished when applied to complex manipulation tasks \cite{lystbaek2024hands, Wagner24, yu2021gaze}. For example, recent studies on object movement \cite{wagner2024gaze} and asymmetric bimanual manipulation \cite{lystbaek2024hands} found that while indirect eye-hand gestures reduced physical effort compared to direct gestures, they did not improve overall performance. Gaze was mainly beneficial for initial selection, while manipulation phases were slower—likely due to limited proprioceptive feedback and the lack of hand preshaping cues \cite{lystbaek2024hands}. These findings suggest a potential ceiling of gaze-hand techniques in complex tasks, highlighting the need for more effective methods for manipulating distant objects.

\subsection{Image Plane Interaction Techniques}
Our goal is to explore how distant object manipulation can be more effectively based on the notion of direct gestures. This is closely related to Pierce et al.'s image plane interaction techniques, which treat the 3D scene from the user's perspective as a 2D image plane, enabling users to interact with distant content through direct hand gestures \cite{pierce1997image}. For instance, the HeadCrusher technique allows users to select a distant object by occluding it with a pinching hand. This makes it plausible to follow up with pinch-based rotation, scaling, and translation (RST) gestures. Such methods could complement \GP, especially because indirect gestures typically keep the hand out of the line of sight. However, depth differences between the hand and distant targets introduce parallax effects, which can cause visual misalignment (e.g., doubled or offset fingers) and create ambiguity in the perceived depth and precise selection point. Later work showed Gaze\&Finger to be more efficient for close targets than distant ones due to parallax \cite{wagner2023fitts}. This limitation was also noted in Pierce et al.’s discussion of image plane techniques \cite{pierce1997image}.

 To improve occlusion-based selection, researchers have explored multimodal combinations with gaze. EyeSeeThrough, introduced by Mardanbegi et al., leverages spatially-coupled eye-hand coordination to eliminate explicit mode switching \cite{Mardanbegi19}. Gaze\&Finger extended this idea to selection tasks by aligning the index finger with the user's gaze in view space \cite{Lystbaek22}, achieving performance comparable to \GP \cite{wagner2023fitts}. This approach has also been applied to region selection in AR \cite{shi2023exploring} and finger typing in VR \cite{lystbaek2022exploring}, where it reduced finger movement compared to standard mid-air typing. 
 
 We initially considered such gaze-based image plane techniques as a pathway to direct object manipulation. However, relying on gaze selection for every gesture departs from the directness of hand gestures. It requires users to constantly shift focus to the target, and the need to synchronize gaze and hand gesture timing can lead to the late-trigger problem \cite{Kumar08}. 




\subsection{Hand Teleporting and Object Summoning}

To address the limitations of distant manipulation, one class of techniques focuses on transporting the user's virtual hand to the distant target. An early example was the Scaled-World Grab locomotion variant proposed by Mine et al. \cite{mine1997moving}, where users are virtually transported toward an object through a single grabbing action. Similarly, the Go-Go technique employs a non-linear mapping between the physical and virtual hand, effectively extending reach beyond physical constraints \cite{poupyrev1996go}. This inspired a range of "virtual-hand teleportation" techniques that enable distant interaction without full-body locomotion \cite{deng2017understanding, jeong2023gazehand, poupyrev1996go, bowman1997evaluation, Zhou2024Reflected}.

An alternative approach tackles the challenge from the opposite direction: by bringing a representation of the distant object (also known as near-field metaphor, proxy, or replica) into the user's reachable space. A seminal work is Stoakley et al.'s World in Miniature \cite{WIM}, where a handheld mini-world represents the entire virtual scene, allowing users to indirectly manipulate the objects in the scene. Bringing objects closer enables users to manipulate them directly while benefiting from proprioception, stereopsis,  head-motion parallax, and  improving manipulation accuracy \cite{mine1997moving}. 

\revised{To better make this concept scale, subsequent research focused on interaction techniques to trigger the proxy creation on the fly for any given virtual context. From an input-theoretical point of view, These can be classified into two categories. First, using a dedicated, additional input command to trigger the proxy creation. For instance, specific gestures (e.g., Poros \cite{Poros_Pohl_Lilija_McIntosh_Hornbæk_2021}) or occlusion selection and bimanual input (e.g., VoodooDolls \cite{VoodooDolls_Pierce_Stearns_Pausch_1999}). These allow a clear separation of intent, at the expense of adding an additional step before one can use the proxy, and an additional command to learn for the user. Second, completely replacing the remote control method. For instance, Scaled-World Grab \cite{mine1997moving} warps the selected object and its context to the user's hand at every gesture. However, this takes away the option to fall back to the default remote control paradigm. Our work extends the prior art through exploring a warping technique without new commands due to exploiting eye-hand coordination patterns.
}

\subsection{Direct/Indirect Mode Switching}
Historically, computer systems have relied on distinct direct and indirect input devices, each offering complementary interaction properties \cite{hinckley07}. 
To harness the strengths of both, hybrid techniques have been developed that allow users to switch between input modes. For example, HybridPointing supports direct pen input on large displays, but transitions to an indirect cursor mode when interacting with a trailing widget \cite{Forlines06}. Similarly, ARCPad extends a touchpad’s relative pointing with an absolute mode, differentiating between tap and drag gestures \cite{McCallum09}.


The multimodal combination of eye-tracking and direct input devices supports both direct and indirect gesture modes modulated by eye-hand coordination. \revised{Some approaches explored explicit mode switching, e.g., FingerSwitches \cite{Pei2024UIMobilityControl} uses \GP micro-gestures to switch UI windows across static, dynamic, and self entities. A distinct category is implicit mode switches without explicit manual input. For instance, Gaze-Shifting \cite{Pfeuffer15} uses implicit modulation based on eye-hand coordination patterns}. This approach defines direct manipulation when manual input falls within a predefined gaze-centric range in a 2D interface, determined by the distance between the gaze point and input position. For instance, a pen's input can seamlessly transition from the default direct drawing mode to indirect menu operation when the user's focus shifts to a distant menu. Such a co-existence of direct and indirect input modes at the granularity of each input command minimizes mode-switching costs \cite{Fleischhauer23}. This principle has been extended to 3D user interfaces, using hand-tracking input device and visual-angle-based range definitions, enabling users to shift between direct and indirect modes for each pinch gesture \cite{Pfeuffer17,lystbaek2024hands}, and is a core feature of the Apple Vision Pro's UI. We extend the prior art by considering how the direct-indirect flexibility can be brought to the manipulation of objects at a distance.





\section{\name Interaction Design}
\name is a novel technique to summon proxies of distant objects into reach through exploiting eye-hand coordination patterns. In the following, we detail the design and parameters of the method.

\subsection{Phases of \name Interaction} \label{sec:design-phases}
\begin{figure}
    \centering
    \includegraphics[width=\linewidth]{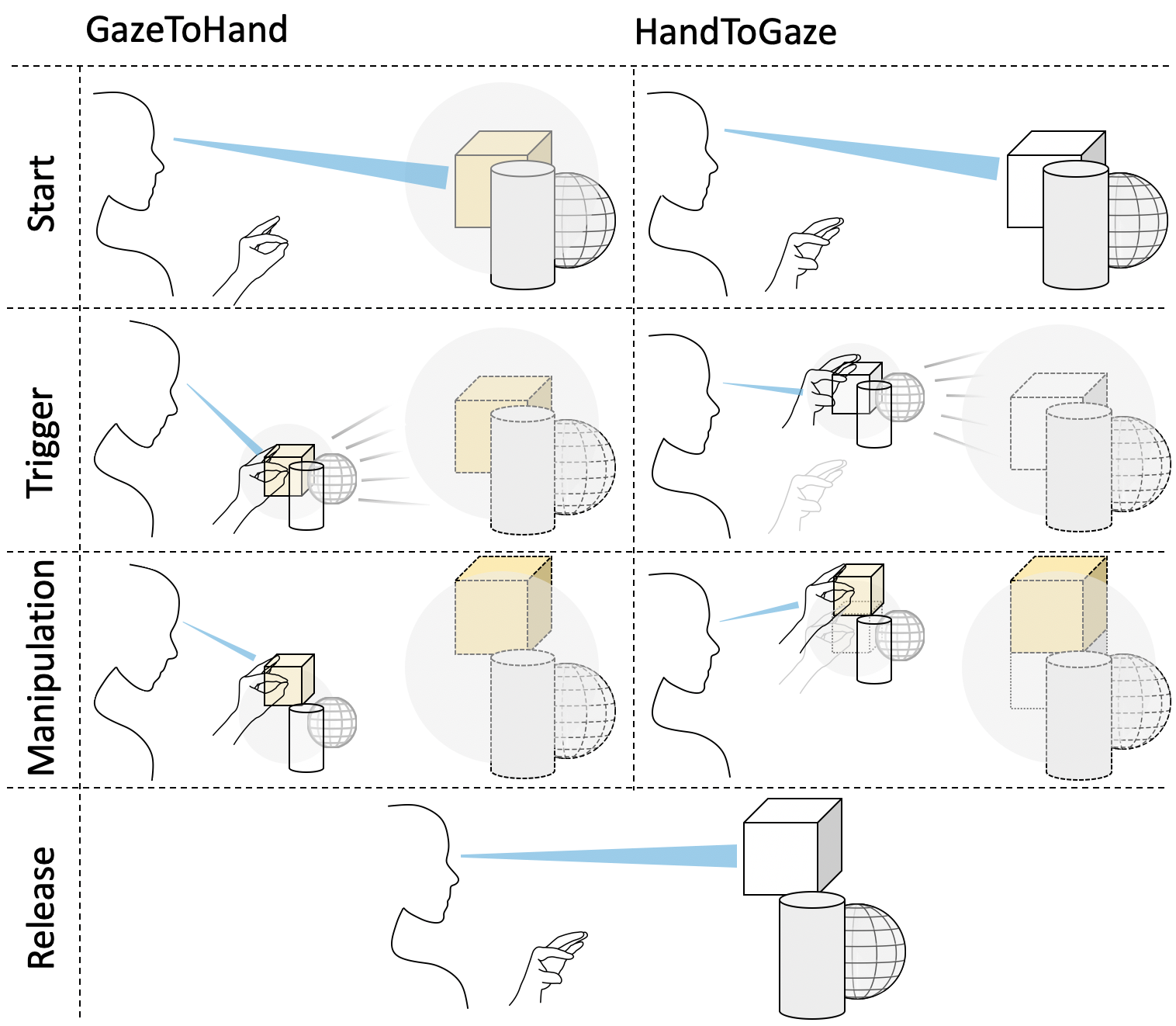}
    \caption{Four-phase illustration of GazeToHand and HandToGaze. The blue triangle indicates the user's gaze. An object is colored yellow when it is being manipulated with indirect or direct gestures. \GH requires users to first preselect a target via \GP, then triggers summoning by directing their gaze toward their pinch hand. Users then apply direct gestures to move the target in near space, and finally release the pinch while moving their gaze or hand away to deactivate summoning and disengage the proxy. In contrast, \HG does not necessitate a \GP pre-selection---users can gaze at a target and move their hand into alignment with gaze to trigger summoning. Users then pinch on the near space object to initiate manipulation.}
    \label{fig:conpect-interaction}
\end{figure}

Our method integrates the act of proxy summoning with subsequent direct gestural operations into a cognitively unified interaction flow. The interaction procedure includes the following states (\autoref{fig:conpect-interaction}): 
\begin{enumerate}
\item \textbf{Start}: A target is identified based on gaze direction and fixation. 
    \item \textbf{Trigger}: Summoning proxies of the target and its context to the user's hand.
    \item \textbf{Manipulation}: Then, users directly operate on the proxy of target or of other objects within the context.
    \item \textbf{Release}: The target and its context returns to the far space with the results of the direct interaction once the trigger condition is no long maintained. 
\end{enumerate}


For \textbf{Trigger}, we adopt a simple eye-hand coordination pattern that avoids interfering with indirect gestures in far space. \GP is typically used with the hand held ergonomically away from the gaze direction, \revised{e.g., near waist level, aligning with prior observations \cite{lystbaek2024hands}}. This makes gaze-hand alignment--either looking at the hand or bringing it into the line of sight--an expressive and yet unused eye-hand coordination pattern during \GP. Albeit prior work has exploited this pattern for various object selection use cases \cite{Lystbaek22,Mardanbegi19,Schweigert19,lystbaek2022exploring}, we re-imagine this pattern for a new purpose: to trigger proxy summoning for direct gestural control. The mechanism includes two simple ways of eye-hand coordination, which we define as two modes in \name's input model:
\begin{itemize}
    \item \GH: Summoning is triggered by users explicitly directing their gaze to the hand after selecting an object with \GP. \GH offers a distinct perspective of the far context by summoning it to the hand (e.g., a top view). 
    \item \HG: \HG is is triggered by bringing the hand to pinch on the line of sight that focuses on a target. \HG creates the visual perception of directly manipulating distant objects while preserving its viewing angle. The near-space proxy can be summoned either at the hand's position or along the gaze line. Summoning to the hand ensures that every pinch can exactly select the gazed target. Summoning along the gaze, while not guaranteeing a successful direct selection upon initial pinch, allows for more flexible hand repositioning to interact with objects other than the initially gazed target within the proxy.
\end{itemize}

For summoning the proxy of a distant object for direct gestural manipulation, its immediate surroundings (\textbf{context}) must also be summoned for contextual reference, following established practice \cite{mine1997moving,WIM,Poros_Pohl_Lilija_McIntosh_Hornbæk_2021}. The contextual summoning preserves relative positioning among objects. It is particularly valuable in the \GH mode, where users can focus entirely on near-space manipulation without needing to visually cross-reference the far-space context.

The \textbf{Manipulation} phase is compatible with but not limited to \GP. Once gaze fixates on a target, the user may either pinch to interact indirectly, or transition to direct interaction via summoning: by pinching and then looking at the hand (\GH), or by performing a spatial alignment of their gaze and hand to trigger \HG. Upon gaze-hand alignment, a proxy of the target and its context is summoned into the near space. This summoned proxy persists as long as the gaze and hand are in proximity within the view plane, regardless of whether the pinch is held. This design allows users to move their hand within the proxy to acquire and manipulate different objects in the context.

When users intend to \textbf{Release} the summoned proxy, they simply move away their gaze or hand to break their spatial alignment.

For \textbf{Trigger} and \textbf{Release}, the angular threshold of gaze-hand alignment for summoning is important. A smaller angle (e.g., 5$\degree$) enables more precise summoning but requires greater effort to align, which has been successfully used for precise selection techniques \cite{lystbaek2022exploring,Lystbaek22}. In contrast, a larger angle (e.g., 30 $\degree$) offers easier activation but might lead to accidental triggering. In our context, we adopt a generous angular threshold as we consider a different task of summoning a proportion of the far space for users to engage in subsequent interaction, where ease of use outweighs precision. Moreover, we apply an even more relaxed threshold for deactivation, reducing the chance of accidentally breaking the spatial alignment when users move their hand around the summoned context.

\autoref{fig:conpect-interaction} illustrates the four phases for both summoning modes of \name. \GH requires users to first select a target via \GP, then triggers summoning by directing their gaze toward their pinch hand. Users then apply direct gestures to move the target in near space, and finally release the pinch while moving their gaze or hand away to deactivate summoning and disengage the proxy. In contrast, \HG does not necessitate a \GP pre-selection, as users can gaze at a target and move their hand into alignment with the gaze to trigger summoning. The user then pinches on a near-space object to initiate manipulation. The release phase is identical between both modes. 

\begin{figure}
    \centering
    \includegraphics[width=\linewidth]{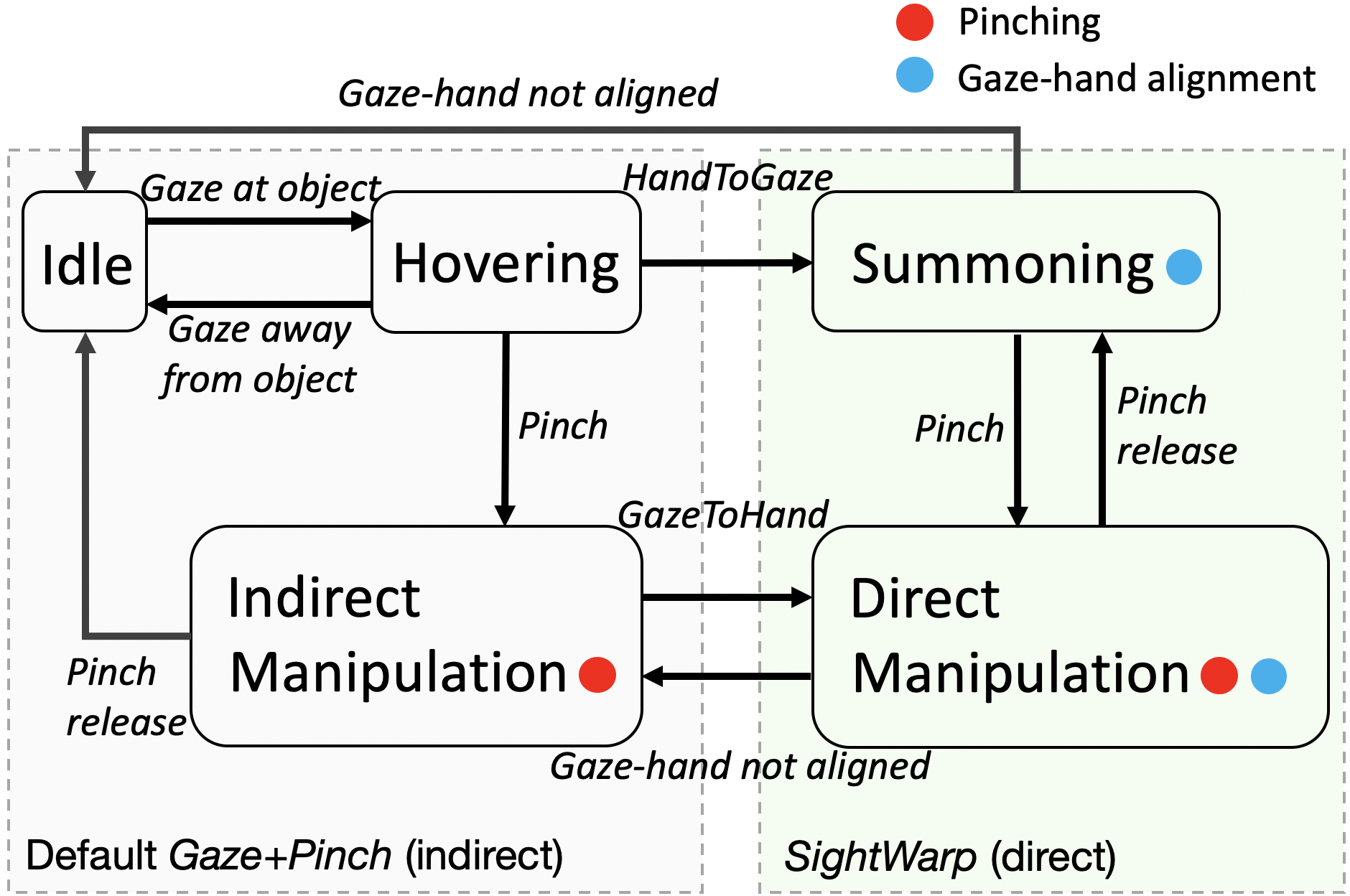}
    \caption{Five-state model illustrating transitions among the three modes: default \GP (left dashed box) and the two \name variants (right dashed box). Colored dots (red and blue) indicate conditions needed to be maintained to remain in a given state. The \textit{Summoning} state denotes that a proxy is summoned but not yet manipulated, while a pinch transitions to the \textit{Direct Manipulation} state for the proxy.}
    \label{fig:concept-state-diagram}
\end{figure}

\subsection{Input State Model}
\autoref{fig:conpect-interaction} illustrates the steps  of using each summoning mode and \autoref{fig:concept-state-diagram} presents  the transitions among five states in a system where \GP, \GH, and \HG modes co-exist. Beyond the existing three states of \GP (\textit{Idle}, \textit{Hovering}, \textit{Indirect Manipulation}), the \HG pathway introduces a transition from \textit{Hovering} to \textit{Summoning}, enabling subsequent \textit{Direct Manipulation} with a pinch. In contrast, \GH triggers direct proxy manipulation from the \textit{Indirect Manipulation} state. Overall, the system provides two bidirectional pathways for switching between indirect and direct modes, featuring our design goal of a fluid, always-available transitioning mechanism. 


\subsection{Design Considerations}



Both the far-space original and near-space proxy contexts are visualized as semi-transparent spheres, each centered on the corresponding target object. \revised{We choose a spherical shape because it provides uniform coverage of surrounding space in all directions}, helping perceive and adjust the scope of both contexts, akin to the design of Poros \cite{Poros_Pohl_Lilija_McIntosh_Hornbæk_2021}.

Upon selection via \GP, a semi-transparent context sphere appears around a selected object. \revised{Any object intersecting this sphere is included in the context to be summoned}. A proxy sphere is summoned into the near space upon triggering gaze-hand alignment. To reduce visual clutter, portions of contextual objects extending beyond the near-space sphere bounds are cropped. A larger far-space context encompasses a wider range of spatial references, while a smaller one captures only the immediate surroundings. \revised{By default, the initial diameter of the far-space context is set to twice the size of the target's bounding box.}


\revised{For the near-space proxy, its size influences the Control-Display (CD) ratio}. Indirect input techniques such as mouse, hand ray, and \GP commonly employ a CD ratio to map hand movement to object translation to alleviate physical effort and arm fatigue when moving objects over large distances. A common practice for \GP distant manipulation is to use a visual angle-based CD ratio, where the object moves across the same angular distance in the user's view as the hand \cite{lystbaek2024hands, Wagner24}. This is achieved by matching the translation distance in visual angle between the target in far space and its proxy in near space. 

\revised{We extend this principle to the \HG mode, which preserves the original viewing angle. Thus, it naturally benefits from the same visual-angle-based CD ratio strategy} by scaling the proxy to match the visual size of the original context. In contrast, the \GH mode is not constrained by visual angle consistency, offering greater flexibility to adjust CD gain for optimising manipulation precision or efficiency, depending on the use case. The CD gain can be adjusted by resizing the near-space context. Users can modify the size of both the far and near context spheres via an arc-shaped handle at the upper-left location of each sphere.

\section{User Study}

This user study investigates the trade-off between the cost of transitioning from indirect to direct manipulation and the potential performance gains enabled by \name. \revised{Existing paradigms for distant object interaction, such as \GP, offer efficient selection but lack the benefits of direct, hands-on manipulation. Our technique, \name, is designed to complement \GP by providing distinct modes for direct gestural interaction. However}, this design introduces a necessary transition cost. \revised{To evaluate this trade-off, we compare simplified variants of \name's summoning mechanisms against a \GP baseline on a 6DOF docking task.} Our research questions (RQs) are as follows:

\textbf{\textsc{RQ1}: How does user performance with \HG and \GH compare to \GP?} For both \name techniques, they are different in the way of triggering (move your gaze, or move your hand), and in the perspective of summoned content (holds perspective, vs. provides a new perspective). What is their effect on the user's performance and experience, and how do they compare against the \GP baseline?

\textbf{\textsc{RQ2}: How does task complexity affect the user's performance with \name?} Given the cost of the context switch, it is unclear which point of task complexity will it become beneficial to use direct gestures in near space, over indirect gestures in far space. We investigate task completion time, manipulation time, clutches and errors across two object sizes and rotation difficulties.

\textbf{\textsc{RQ3}: How well can users perform the initial summoning?} Users need to change their focus distance and change the interaction paradigm from indirect to direct gesture. How do people manage the context shift, and what are potential costs to this context change? 

We conducted a within-subject study. We employed a 3$\times$2$\times$2 factorial design of the following independent variables, with condition order counterbalanced across participants:
\begin{itemize}
    \item Techniques: \GP (Baseline), \GH, \HG
    \item Rotation Magnitude (between initial/target rotation): 45$\degree$, 90$\degree$
    \item Object Size (in visual angle): 7.5$\degree$, 12.5$\degree$
\end{itemize}


\begin{figure}
    \centering
    \includegraphics[width=\linewidth]{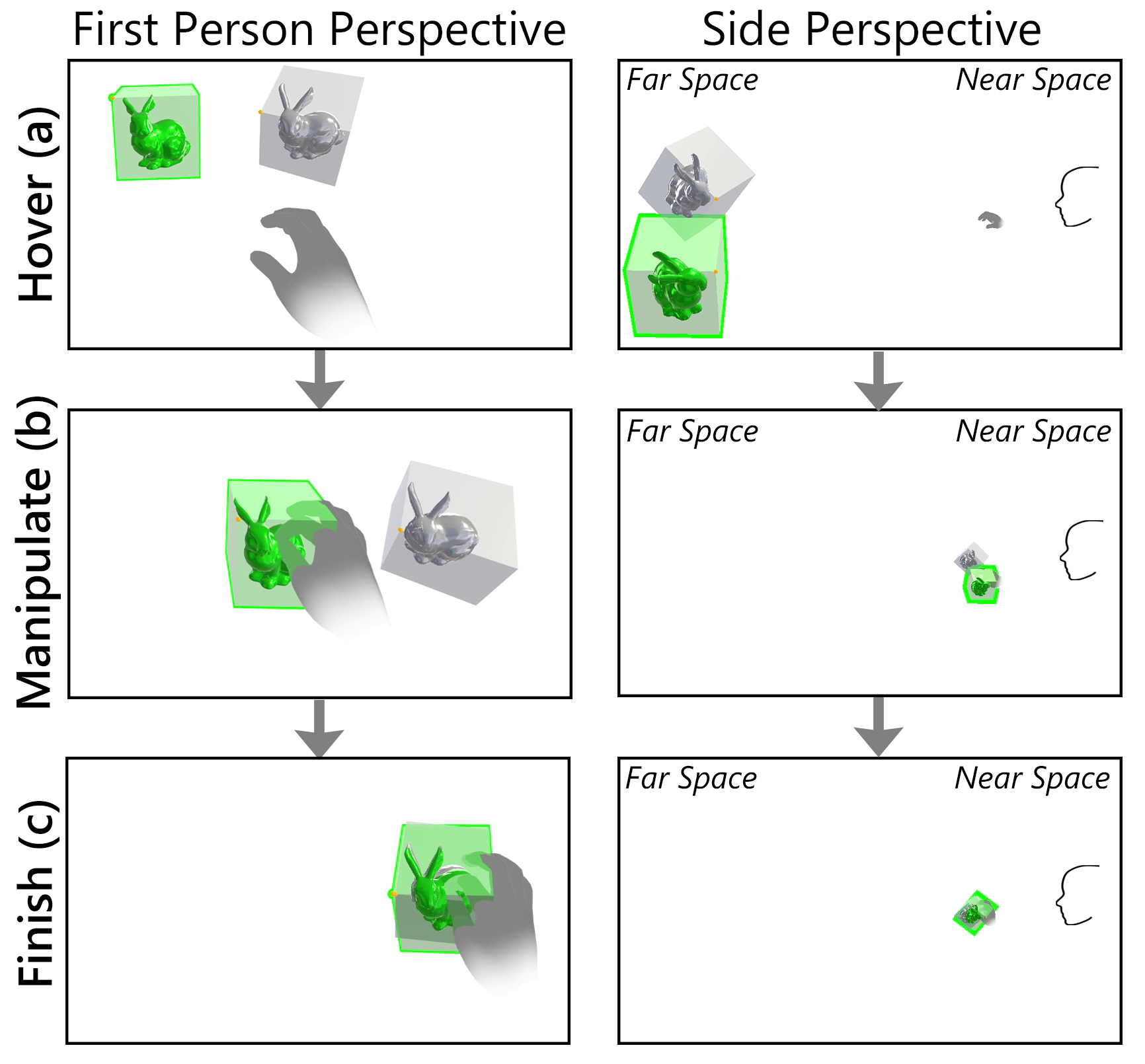}
    \caption{Trial sequence for the two summoning conditions (\GH and \HG). The left column shows the first-person perspective, and the right column shows the side view. (a) The user hovers over the object using gaze. (b) The object and the target are warped to the user's hand as a direct manipulation proxy. (c) The trial is completed once both translational and rotational thresholds are met.}
    \label{fig:TrialSequence}
\end{figure}

\subsection{Task Design}
The task is a 6DOF docking task, where participants manipulate a 3D object to match the position and orientation of a target object \cite{10.1145/3411764.3445193}. \revised{This task is chosen because it is the state-of-the-art evaluation method for 3D manipulation, which \name enables. While \name is also applicable to other tasks, such as selection and multi-step workflows as explored in \autoref{sec:applications}, evaluation of these scenarios is out of scope.}

The manipulable object is a semi-transparent green cube that encloses an opaque Stanford bunny \cite{yu2021gaze}, with a size of either 7.5$\degree$or 12.5$\degree$ in visual angle (26.2 cm or 43.8 cm wide at 2 m viewing distance). The target object is identical in shape and size, but rendered in gray and not interactive. 

Each trial begins when the object and its target appear, and ends when their positional offset falls below 20\% of their visual angle size, and their orientational difference is within 15$\degree$ in quaternion angle. \revised{A three-step trial sequence for the two summoning conditions is illustrated in \autoref{fig:TrialSequence}}. To avoid accidental completions due to brief or unstable alignments, participants need to maintain these completion conditions for 300 ms. Clutching is allowed.

At the beginning of each trial, the manipulatable cube appears 2 meters in front of the participant, front facing and chest-aligned, as approximated from the HMD position. The target object is positioned at an offset of twice the object's visual size along one of four displacement directions (+X, -X, +Z, -Z), and rotated by either 45° or 90° around one of three randomly selected axis pairs ($\pm$X$\pm$Y, $\pm$Y$\pm$Z, $\pm$X$\pm$Z). \revised{The Y-axis was omitted from positional displacement to reduce study complexity.} 

Each block (a unique combination of Technique, Rotation Magnitude, and Object Size) included 12 trials, covering all 4 displacement directions crossed with 3 randomly assigned rotation axes. The trial order within each block is randomized. In total, each participant completed 3 Techniques $\times$ 2 Rotation Magnitudes $\times$ 2 Object Sizes  $\times$ 12 combinations = 144 trials.






\subsection{Procedure}
Participants were first briefed on the study and completed consent and demographics forms. Before each condition, they watch an instructional video demonstrating the respective technique. They then wore the headset and performed fit adjustment and eye-tracking calibration. Participants remained seated in a static, non-swiveling chair throughout the study, allowing only upper-body movement. 
For each condition, participants began with a hands-on training session in a task-free environment, practicing the technique until they felt comfortable. They then completed four blocks of 12 docking trials, with varying object sizes or rotation magnitudes. Breaks were allowed between blocks, and the next block was initiated upon confirmation with the researcher. Participants were instructed to perform as fast as possible. 
After each condition, participants completed a post-condition questionnaire and repeated headset fitting and gaze calibration. At the end of the session, they completed a post-study questionnaire. 

\subsection{Evaluation Metrics}
\revised{We collected the following metrics to assess task performance, perceived workload, and user experience.}

\begin{itemize}
    \item \TCT: time from object appearance to trial completion, i.e., meeting the accuracy thresholds.
    \item \AT: time from object appearance to the first pinch.
    \item \FMD: duration of the first pinch.
    \item \CC: number of clutch gestures.
    \item \FGC: number of pinch gestures that had not no effect on the object (i.e., failed grabs).
    \item \TDM: total hand travel distance while pinching.
    \item \RDM: total hand rotation while pinching.
    \item After each condition, participants completed the NASA TLX questionnaire \cite{HART1988139} to report perceived workload.
    \item After completing all trials, participants rated their overall experience for each technique and provided written feedback.
\end{itemize}

\subsection{Apparatus and Implementation}
The study was implemented in Unity (2022.3.19f1) for the Meta Quest Pro (90 Hz display, 30 Hz eye tracker), using the Meta XR All-in-One SDK (v74.0.1). Hand-tracking data was smoothed using a 1€ Filter \cite{1EUROfilter}, and pinch gestures were detected with a relaxed confidence threshold for easier acquisition. During manipulation, a green outline highlights the selected object.

To ensure consistency across all techniques, both summoning and manipulation were initiated with a pinch gesture. In the \HG and \GH conditions, summoning occurs at the moment of pinching, enabling immediate manipulation in near space within the same gesture session. To encourage near-space interaction and avoid visual clutter, far-space objects were removed immediately upon summoning.

For the \HG condition, we set the angular threshold for gaze-hand alignment to 25$\degree$, \revised{determined through pilot testing to balance ease of triggering, as discussed in \autoref{sec:design-phases}}. Piloting also revealed that users sometimes brought their hand too close to their face, causing the summoned object to appear uncomfortably near. To mitigate this, we added a depth constraint: the hand had to be within 0.3–0.5 m from the user to trigger summoning. For exiting, the threshold was slightly relaxed to 30$\degree$, and the valid depth range was extended to 0.25–0.65 m. These buffers help prevent unintended exits, such as slipping out of range during mid-pinch.

To ensure comparability across techniques, object movement was computed based on visual angle rather than direct 1:1 hand displacement. Specifically, movement was scaled by the ratio between the distance from the far-space object to the user’s eyes and the distance from the hand to the eyes, ensuring that positional manipulation remains consistent in terms of angular displacement across all techniques. \revised{For \GH and \HG, summoned objects were scaled down based on this same control-display ratio, preserving their original visual size.}

\subsection{Participants}
12 participants (4 female, 8 male) took part from the local area, primarily university students. Participants ranged in age from 22 to 35 (\textsc{M = 26.91, SD = 4.09}). All were right-handed or ambidextrous; 4 wore glasses and 2 wore contact lenses. On a 5-point scale, participants reported little to moderate experience with VR/AR (\textsc{M = 2.67 SD = 1.17}), 3D hand gestures (\textsc{M = 2.41 SD = 1.32}), and gaze input (\textsc{M = 2.41 SD = 1.38}).

\subsection{Results}

For task performance data, we applied the Aligned Rank Transform (ART) to address deviations from normality~\cite{Wobbrock2011Aligned}. Next, we performed a repeated measures ANOVA with performance data, and post hoc pairwise comparisons (Holm-Bonferroni corrected).
For NASA-TLX and preference ratings, we performed a Friedman’s Test and found no significant results. We plot results for measures that yielded significant results regarding \TECH in \autoref{fig:Acquisition Time}-\ref{fig:Manipulation Movement} while only reporting main effects of \RM and \OS in text. Statistical significance is shown as * (p < .05), ** (p < .01), and *** (p < .001). Error bars indicate standard deviation.

\subsubsection{\TCT (\autoref{fig:Acquisition Time})}
We found significant effects in \TECH ($F_{2,121} = 20.09, p < .001$), \RM ($F_{1, 121} = 193.78, p < .001$), \OS ($F_{1, 121} = 8.87, p < .01$), and \textsc{Rotation$\times$Size} ($F_{1,121} = 4.25, p < .05$). 
Post hoc comparisons revealed that \GP was slower than both \GH ($p < .001$) and \HG ($p < .001$). Performance was also significantly slower for the larger \RM ($p < .001$) and for the smaller \OS ($p < .001$).

\subsubsection{\AT (\autoref{fig:Acquisition Time})}
\revised{While we did not find significant effect, we plot the data in \autoref{fig:Acquisition Time}.}

\begin{figure}[t]
    \centering
    \includegraphics[width=1\columnwidth]{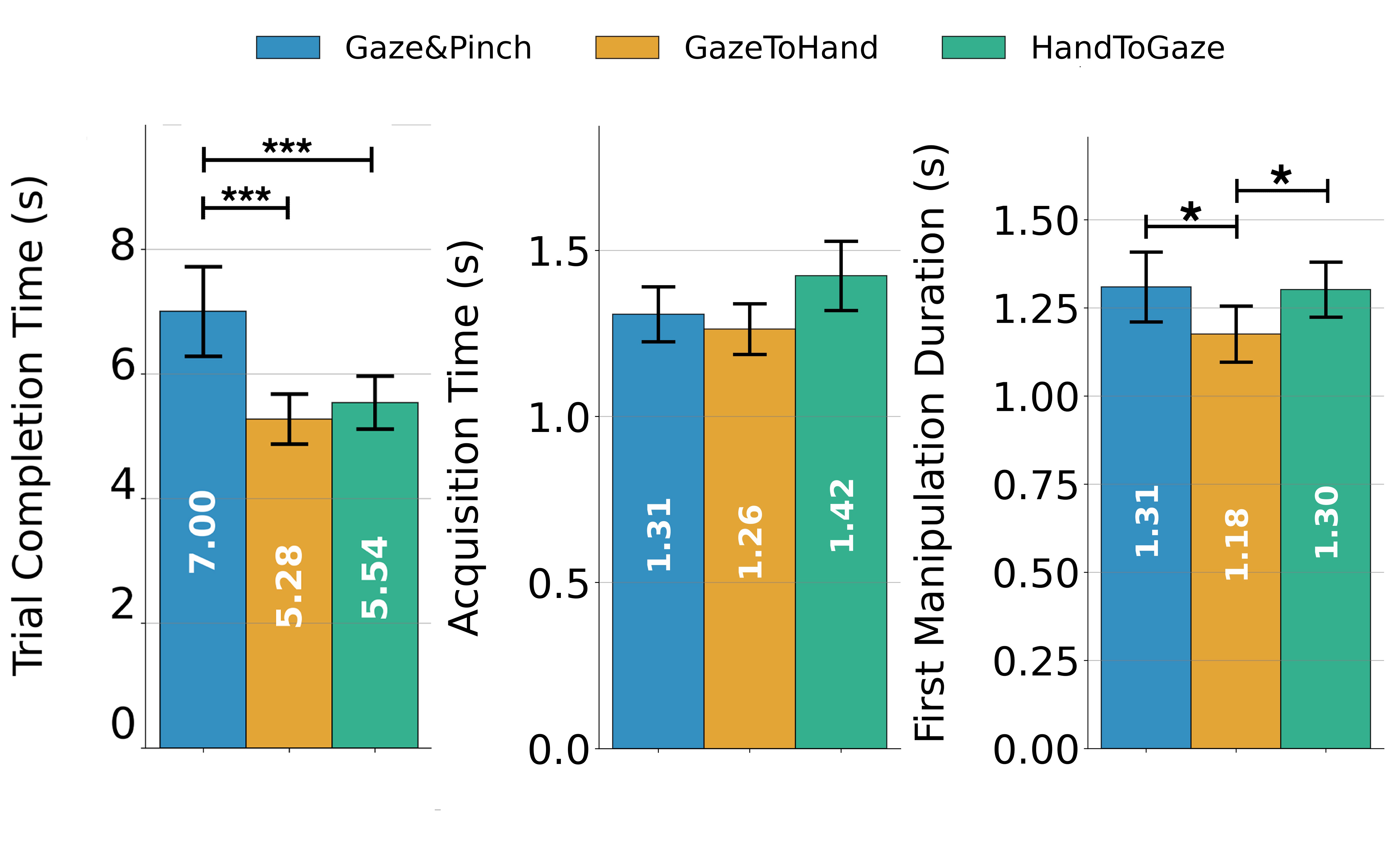}
    \caption{Mean \TCT, \AT, and \FMD.}
    \label{fig:Acquisition Time}
\end{figure}

\subsubsection{\FMD (\autoref{fig:Acquisition Time})}
We found significant effects in \TECH ($F_{2,121} = 4.73, p < .05$) and in \RM ($F_{1, 121} = 22.61, p < .001$). Post hoc comparisons revealed that the \FMD of \GH was shorter than for \GP ($p < .05$) and \HG ($p < .05$). Additionally, the duration was shorter for the smaller \RM ($p < .001$).


\captionsetup{aboveskip=-8pt, belowskip=-6pt}

\subsubsection{\CC (\autoref{fig:Number of Error Manipulations})}
We found significant effects in \TECH ($F_{2,121} = 14.10, p < .001$), \RM ($F_{1, 121} = 335.05, p < .001$), \OS ($F_{1, 121} = 4.57, p < .05$), and \textsc{\RM$\times$\OS} ($F_{1,121} = 6.12, p < .05$). \GP induced more clutches than both \GH ($p < .001$) and \HG ($p < .001$). Larger \RM ($p < .001$) and smaller \OS ($p < .05$) induced more clutches. 

\begin{figure}[t]
    \centering
    \includegraphics[width=.72\columnwidth]{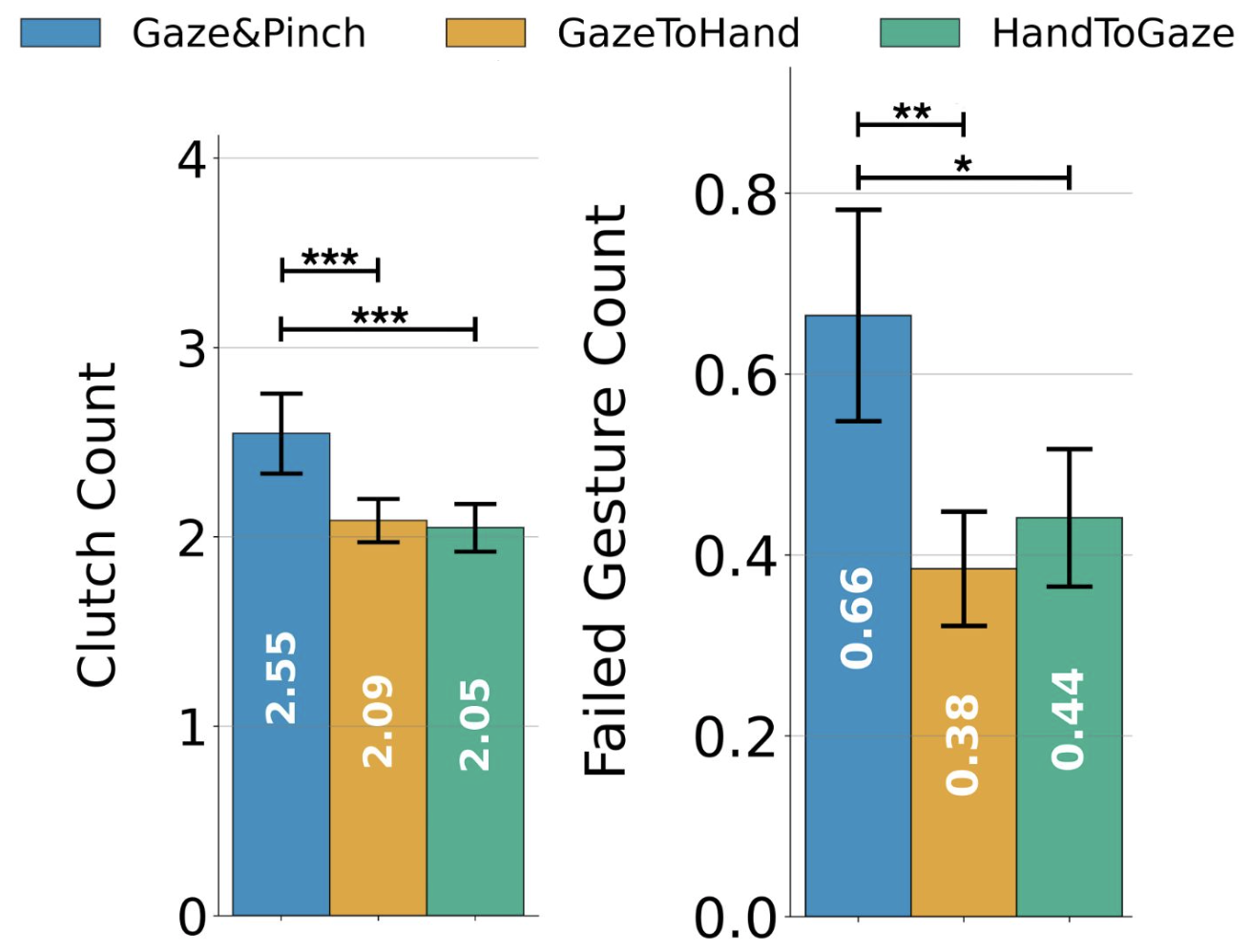}
    \caption{Mean \CC and \FGC.}
    \label{fig:Number of Error Manipulations}
\end{figure}

\subsubsection{\FGC (\autoref{fig:Number of Error Manipulations})}
We found significant effect in \TECH ($F_{2,121} = 10.77, p < .001$). Post hoc comparisons revealed that \GP induced more errors than both \GH ($p < .01$) and \HG ($p < .05$).

\captionsetup{aboveskip=2pt, belowskip=-6pt}

\subsubsection{\TDM (\autoref{fig:Manipulation Movement})}
We found significant effects in \TECH ($F_{2,121} = 4.49, p < .05$), \RM ($F_{1, 121} = 213.39, p < .001$), and \OS ($F_{1, 121} = 21.92, p < .001$). We find that \GH induced less hand translation than \GP ($p < .01$). Respectively, the larger \RM ($p < .001$) and \OS ($p < .001$) induced more hand translation.

\begin{figure}[t]
    \centering
    \includegraphics[width=0.8\columnwidth]{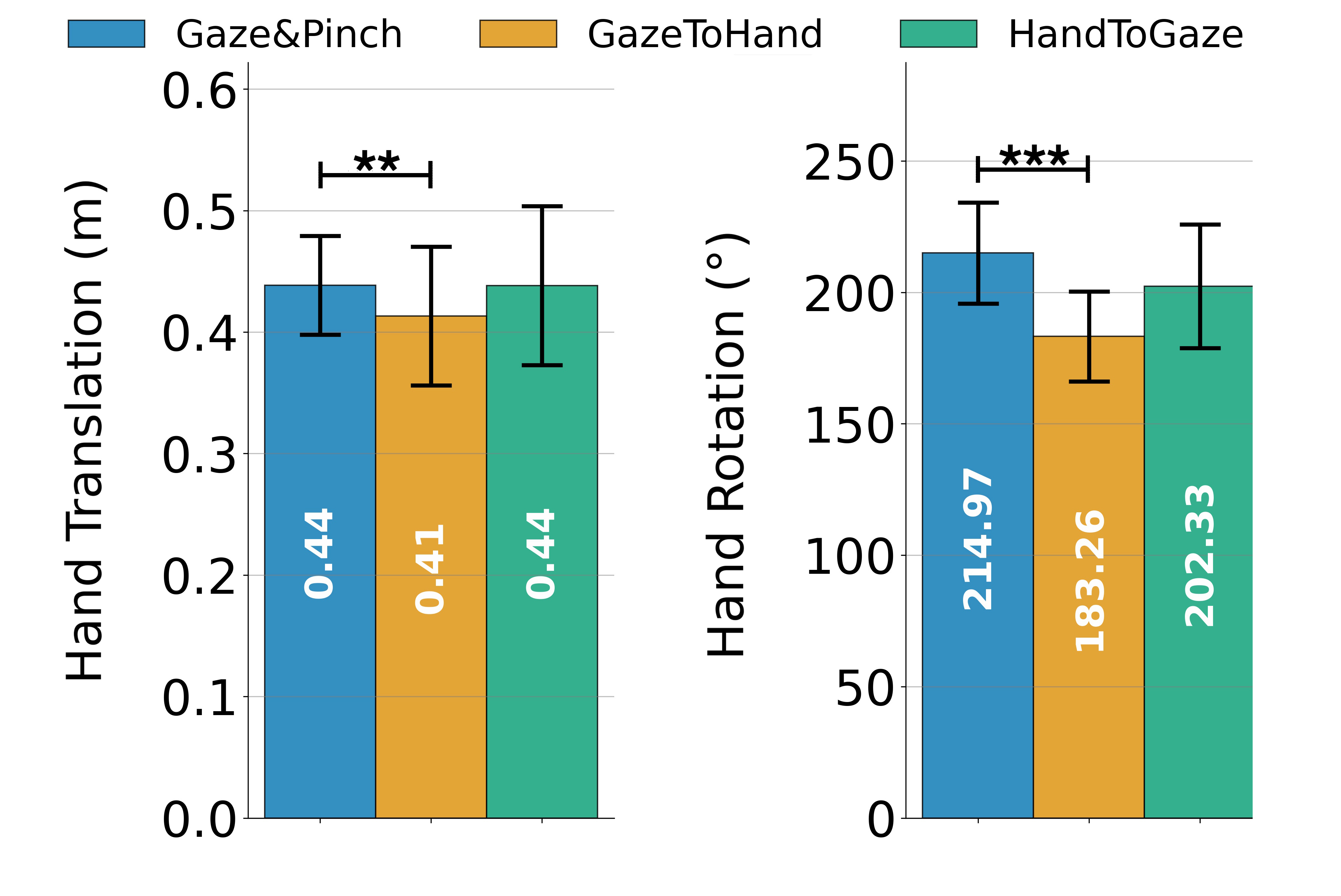}
    \caption{Mean \TDM and \RDM.}
    \label{fig:Manipulation Movement}
\end{figure}

\subsubsection{\RDM (\autoref{fig:Manipulation Movement})}
We found significant effects in \TECH ($F_{2,121} = 9.45, p < .001$) and \RM ($F_{1, 121} = 466.10, p < .001$). Post hoc comparisons show that \GH induced less hand rotation compared to \GP ($p < .001$). Respectively, the larger \RM ($p < .001$) induced more hand rotation.

\captionsetup{aboveskip=2pt, belowskip=-10pt}

\subsubsection{User Feedback}
Participants generally found \GP natural (2 participants) and offering good control of the object (4), but also fatiguing (3). In contrast, \GH was perceived as less tiring (3) and easier to use (7). Two users favored the hand alignment posture in \HG. However, four participants reported eye strain when refocusing on near objects across both \GH and \HG. 

\revised{Overall, user feedback focused on perceived control and arm fatigue, with \GH and \HG generally favored over \GP in these aspects, despite an increase in eye fatigue. }


    

\subsection{Discussion}


Regarding \textbf{RQ1}, our results show that both \GH and \HG significantly outperformed the baseline \GP across all task performance measures, including \TCT, \CC, and \FGC. Participants completed the docking task faster, with fewer clutch gestures to re-orient the hand and fewer failed attempts to grab and manipulate the object using \GH and \HG. These findings suggest that both techniques afford better spatial manipulation by supporting direct manipulation. 


Besides the main effects of \TECH, we observed main effects of \RM in \TCT and \CC, where 90$\degree$ rotations consistently yielded worse performance than 45$\degree$. 
These findings address \textsc{\textbf{RQ2}}, indicating that the performance benefits of direct manipulation \revised{in near space using \HG and \GH is consistent over \GP.}


\textbf{RQ3}  investigated whether adapting to the depth view change between the original object and the summoned proxy would cause temporal overhead for acquisition and overall manipulation. 
\revised{Because we found no significant difference in \AT among the three technique conditions, this suggests that users were able to handle the initial summoning and context switch with minimal effort. Although \HG involves an additional phase of explicit gaze-hand alignment compared to \GH}, we cannot conclude that this overhead led to inferior performance, as there was no significant difference in \TCT between them, likely due to their distinct eye-hand coordination patterns. Furthermore, our \FMD measure showed that the initial manipulation was significantly shorter with \GH than the other two techniques. This suggests that participants might have adopted a strategy with \GH: they would rapidly direct their gaze to their hand after summoning the object, then release the pinch to plan subsequent manipulations. Many participants reported that \GH was "easy and effortless to use", which may specifically refer to the summoning action. In contrast, \HG might offer subtle benefits for planning both the acquisition and initial manipulation, thanks to preserving of the object's visual angle from the user's perspective.

\section{Application Scenarios} \label{sec:applications}
\begin{figure}[t]
   \centering
   \includegraphics[width=0.8\linewidth]{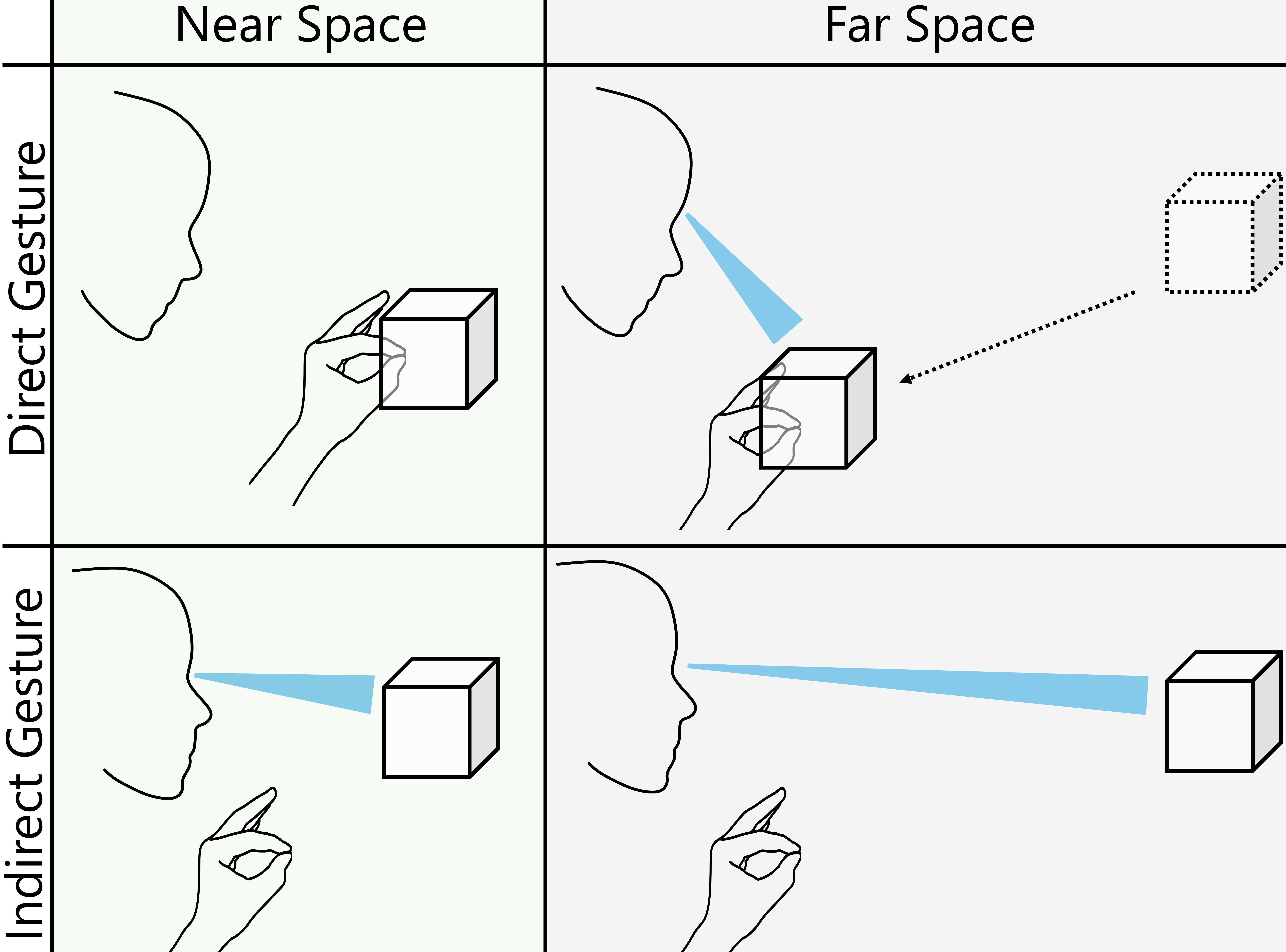}
   \caption{Co-existing modes: Eye-Hand XR UIs support flexible switching of gesture modes, to reap the benefits of direct and indirect inputs for objects across near and far spaces.}
   \label{taxonomy}
\end{figure}

\revised{This section demonstrates the applicability of \name across various XR use cases, including multi-step workflows, by showcasing how its different parameters and design choices can be modulated to facilitate broader near-far interaction paradigms.
In principle, \name expands near-far interaction of XR UIs by incorporating the paradigm of direct-indirect interaction for distant objects.  \autoref{taxonomy} illustrates how all four parts can be integrated in the same UI, through mode-switching mechanisms:
\begin{description}
    \item [Direct \& Near] By default, direct manipulation is active when a hand intersects with a virtual object.
    \item [Indirect \& Far] \GP is active when interacting with a distant object from a convenient hand position.
    \item [Indirect \& Near] Users can interact with a nearby object if the hand is offset from the gaze-selected object, for occlusion-free \cite{hinckley07}, flexible-CD-gain \cite{casiez08} and low-effort interaction  \cite{Wagner24}. 
    \item [Direct \& Far] When looking at a faraway object and performing \GH or \HG, the user  summons object proxies in near space to manipulate them directly.
\end{description}}

\captionsetup{aboveskip=4pt, belowskip=-6pt}


An example for a generic usage is to utilise \HG as  a "zooming" metaphor  (\autoref{appteaser}c--d). 
Direct gestural interaction such as tapping with 2D UI in XR benefits scenarios when gaze pointing suffers from crowded interfaces, and also affords natural interaction similar to physical touch interfaces. Our approach extends these benefits of direct manipulation to distant UIs. While the interaction of \HG is similar to previous work \cite{lystbaek2022exploring, wagner2023fitts, Lystbaek22} based on image plane techniques \cite{pierce1997image}, our summoning mechanism brings UI elements closer in front of the user's hand, which naturally resolves the parallax issue identified in the prior studies. \autoref{appteaser}c--d demonstrates a summoned toggle via a direct tapping gesture.

A key issue in working in 3D is occlusion, and the need for seeing a 3D model from different perspectives, to which \name provides an elegant solution. For instance, object movement in 3D can be challenging as depth change from the user's forward perspective is difficult to perceive. With \GH, users can quickly switch perspectives to perform this task more efficiently. 

We show several application examples for 3D design. We support the interaction with a visual feedback in form of a sphere to indicate what part of the world will be summoned and the boundary of a near-space proxy. Both far and near context spheres have an arc-shaped handle at their upper-left location used to adjust the context size. Since the dominant hand is occupied for object manipulation tasks, the non-dominant hand is used to interact with the scaling handle. To maintain interaction consistency within each space, the far context handle is selected using \GP (indirect gesture) while the near one is grabbed using direct pinch. After selecting a handle, the user moves their non-dominant hand horizontally to change the distance between their two hands, resulting in scaling up or down the corresponding context proportionally.

\captionsetup{aboveskip=2pt, belowskip=-6pt}
\begin{figure*}
    \centering
    \includegraphics[width=\linewidth]{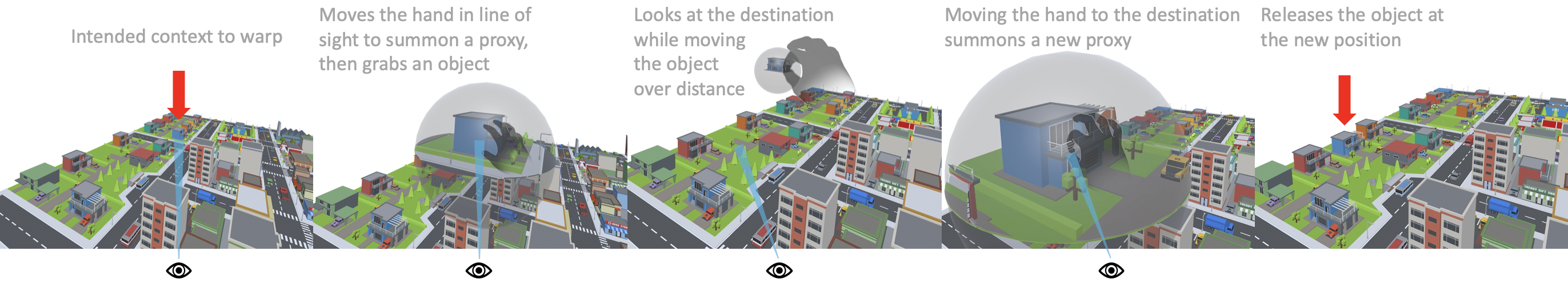}
    \caption{The user performs a \HG drag-and-drop to move the blue building from a starting position to a destination. The eye icon and blue triangle indicate the user's gaze.}
    \label{fig:Applications-drag-n-drop}
\end{figure*}

\captionsetup{aboveskip=4pt, belowskip=-8pt}
\begin{figure}[t]
   \centering
   \includegraphics[width=1\linewidth]{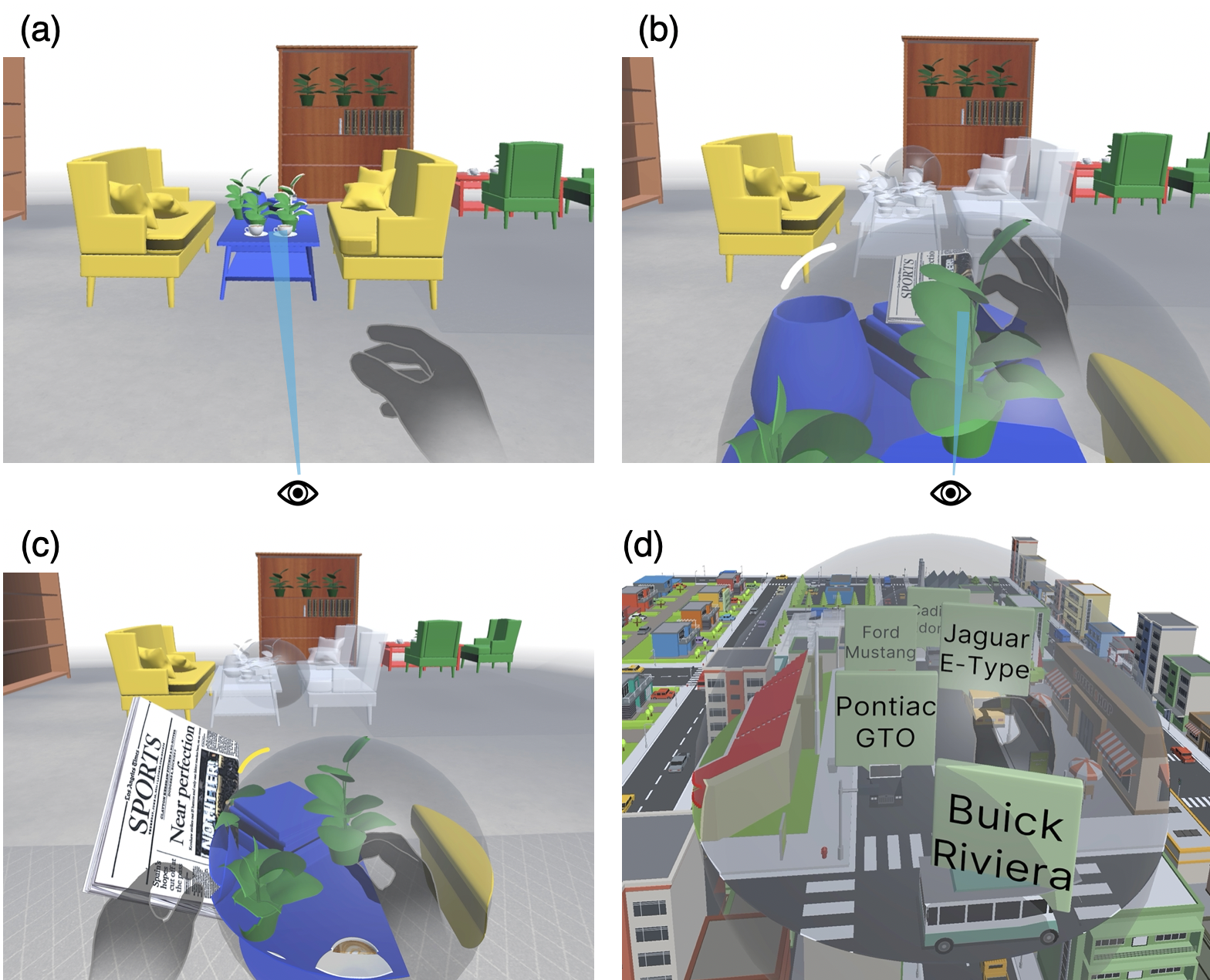}
   \caption{(a)--(b) Using \GH, the user can locate the newspaper that was previously occluded and too small in far space. (c) The user directly acquires the newspaper using the other hand from the summoned near context for further use. (d) Detailed information tags are only rendered when vehicles are summoned into near space. The eye icon and blue triangle indicate the user's gaze.}
   \label{fig:Applications-remaining}
\end{figure}

\subsection{Cross-Space Drag-and-Drop  (\autoref{fig:Applications-drag-n-drop})}
Both \GH and \HG modes provide a close-up view for precise Drag-and-Drop during the task. A workflow enabled by \HG, for example, is to employ direct manipulation for fine-grained operations and indirect manipulation for rapid long-distance movement--toggled at a glance. 
For example, after selecting an object, users can either move it in near space by looking at the near-space proxy while dragging, or switch to indirect manipulation by looking off toward the destination. Summoning can be reactivated by moving the pinching hand to the destination area and re-aligning it with gaze for precise placement.

Another use case is shown in \autoref{fig:Applications-remaining}c, where users can use the non-dominant hand to retrieve a newspaper from the context proxy and place it into their near space. This affordance also allows users to potentially put a new object, using the non-dominant hand, into a context sphere that is summoned into the near space using the dominant hand. In these contexts, the combined use of both hands functions as a distance grab mechanism, providing efficient shortcuts for drag-and-drop operations. This approach could significantly reduce physical effort and speed up the interaction by eliminating the need to traverse between distant spatial contexts.


\subsection{Details on Demand (\autoref{fig:Applications-remaining}d)}
\name extends prior WIM techniques in their  support for Shneiderman’s well-known mantra: “overview first, zoom and filter, then details on demand” \cite{shneiderman1996mantra} through a rapid on-demand creation of WIMs. For example, in an urban design application, summoning a local context can reveal additional details such as annotations or object tags—information that would otherwise cause visual clutter if displayed in the full overview \cite{Lindlbauer19}. As shown in \autoref{fig:Applications-remaining}d, information tags on top of vehicles only appear when the objects are brought into the near context sphere. 

\subsection{Occluded and Small Objects (\autoref{fig:Applications-remaining}a--b)}
Occluded object selection is a classic challenge for raycasting-based methods especially when users stay in stationary positions and perceive the scene from a single perspective \cite{yu2020fully}. \GH addresses this challenge by enabling a perspective change as users can pinch their hand at any position in space to summon the selected context to that location, therefore revealing previously occluded objects. Additionally, the scalability of the context allows users to zoom in on the near context and interact with objects which would be too small to select at a distance. \autoref{fig:Applications-remaining}a--b demonstrates how a newspaper, which is small in size and occluded by a plant, can be easily accessed after summoning the plant's context to near space. 



\section{Conclusion}
 As XR operating systems such as Google's AndroidXR and Apple's visionOS increasingly adopt multimodal input, our research  informs the design of future UIs that build on the direct-indirect input paradigms and contribute to seamless and efficient user experiences. In particular, our work shows that using eye-hand coordination to trigger proxy summoning and transition to direct manipulation is easy to use and efficient.
While far-space interaction will likely remain dominant in many practical XR scenarios, minimizing the effort required to transition to direct manipulation may make near-field interaction a more viable and attractive option.

Our work presents several limitations and directions for future research. First, further studies are needed to explore how \name can be integrated with other interaction modes in realistic, system-wide use cases. While our study demonstrated performance benefits for 3D object manipulation—a core task in spatial environments—it remains to be seen how \name performs in broader applications, such as \revised{selection tasks and multi-step workflows} illustrated in \autoref{sec:applications}. 
Furthermore, we employed a set of self-tested  parameters, e.g., for entering the \HG and \GH modes, which can be further optimised for generic UI as well as specifically-tailored application needs. \revised{While our approach assumes that gaze-hand alignment is infrequent during indirect gestures based on prior work and common usage patterns, further empirical validation would strengthen this assumption.} \name may also be extended with a wider range of proxy summoning methods. For instance, our eye-hand concept could be combined with prior work that uses only eyes vergence to switch between UI depth layers \cite{Zhang24, Hirzle19}. Finally, as with other summoning techniques \cite{WIM, Poros_Pohl_Lilija_McIntosh_Hornbæk_2021}, there are open challenges to address regarding spatial conflicts when summoned objects overlap with existing ones, which can be potentially exploited for merging near and far context \cite{Poros_Pohl_Lilija_McIntosh_Hornbæk_2021}.

\begin{acks}
\revised{This work was supported by funding from a Google Research Gift award ('Multimodal and Gaze + Gesture Interactions in XR'), the European Research Council (ERC) under the European Union’s Horizon 2020 research and innovation programme (grant no. 101021229 GEMINI), and the Danish National Research Foundation under the Pioneer Centre for AI in Denmark (DNRF grant P1).}
\end{acks}

\bibliographystyle{ACM-Reference-Format} \balance
\input{main.bbl}

\end{document}